\documentclass[preprint,tightenlines,aps,floats,nofootinbib]{revtex4}
  
\usepackage{epsf}
\usepackage{graphics}

\newcommand{\notE}{\ \hbox{{$E$}\kern-.60em\hbox{/}}}
\newcommand{\notp}{\ \hbox{{$p$}\kern-.43em\hbox{/}}}
\def\D0{\mbox{D\O}}

\includegraphics{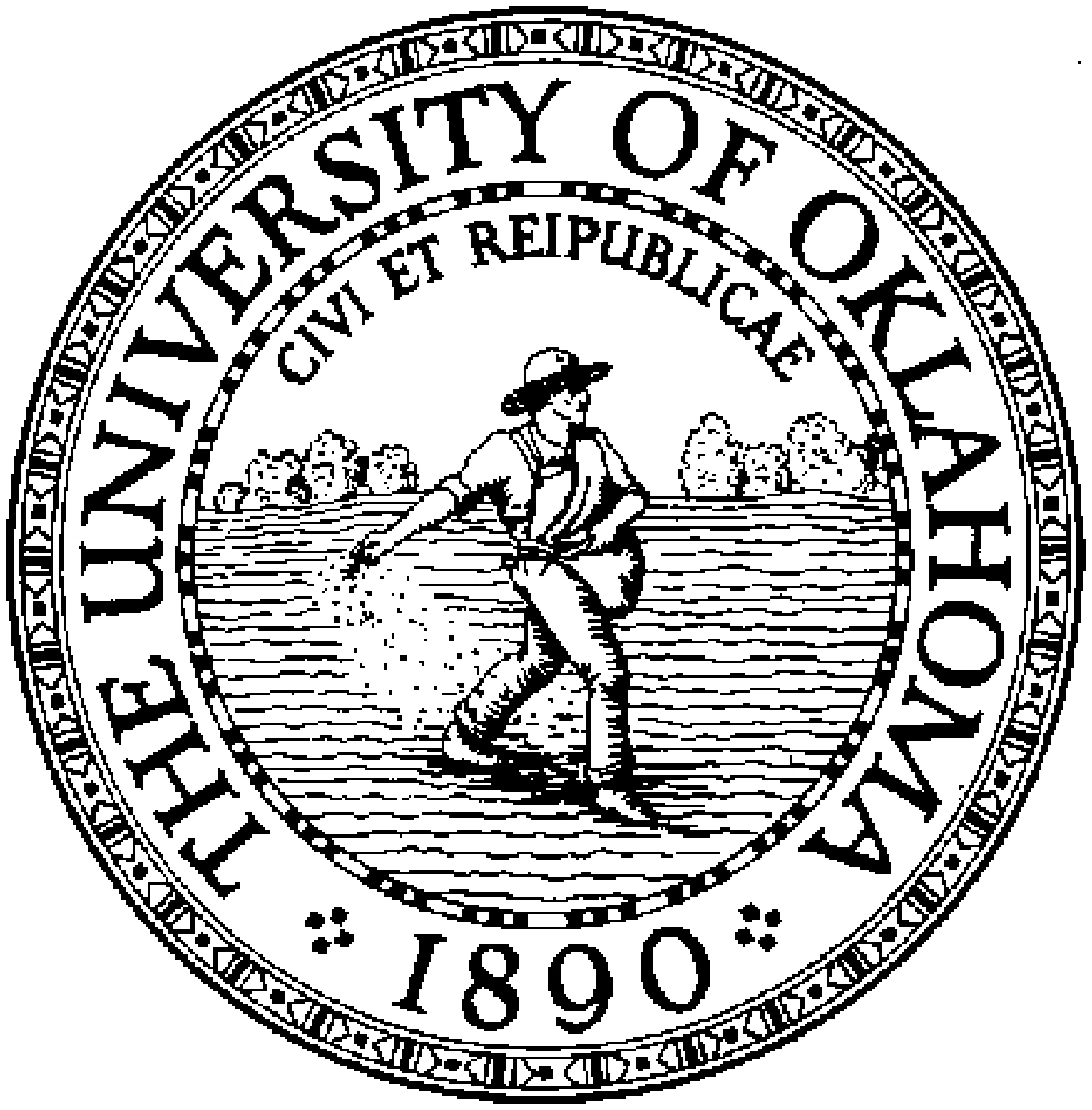}

\preprint{\font\fortssbx=cmssbx10 scaled \magstep2
\hbox to \hsize{
\hskip1.2in 
\hbox{\fortssbx The University of Oklahoma}
\hskip0.2in $\vcenter{
                      \hbox{\bf hep-ph/0610418}
                      \hbox{October 2006}}$ }
}
  
\begin{document}  

  
\title{\vspace*{0.7in}
$B_s \to \mu^+\mu^-$ versus Direct Higgs Searches\footnote{
Presented at the 14th International Conference on Supersymmetry
and the Unification of Fundamental Interactions (SUSY06), 
Irvine, California, USA, 12 - 17 June 2006.}}
 
\author{
Chung Kao\footnote{E-mail address: Kao@physics.ou.edu (C. Kao)} and 
Yili Wang\footnote{E-mail address: Yili@physics.ou.edu (Y. Wang)}}

\affiliation{
Homer L. Dodge Department of Physics and Astronomy \\
University of Oklahoma \\ 
Norman, Oklahoma 73019, USA 
\vspace*{.5in}}

\thispagestyle{empty}

\begin{abstract}

We investigate the prospects for the discovery of neutral Higgs bosons 
with muons by direct searches at the CERN Large Hadron Collider (LHC) 
as well as by indirect searches in the rare decay 
$B_s \to \mu^+\mu^-$ at the Fermilab Tevatron and the LHC.
Promising results have been found for the minimal supersymmetric standard
model, the minimal supergravity (mSUGRA) model, and supergravity
models with non-universal Higgs masses (NUHM SUGRA).
For $\tan\beta \simeq 50$, we find that
(i) the contours for a branching 
fraction of $B(B_s \to \mu^+\mu^-) = 1 \times 10^{-8}$ 
in the parameter space are very close to the $5\sigma$ contours for 
$pp \to b\phi^0 \to b\mu^+\mu^- +X, \phi^0 = h^0, H^0, A^0$ at
the LHC with an integrated luminosity ($L$) of 30 fb$^{-1}$, 
(ii) the regions covered by 
$B(B_s \to \mu^+\mu^-) \ge 5\times 10^{-9}$ and the discovery
region for $b\phi^0 \to b\mu^+\mu^-$ with 300 fb$^{-1}$ are
complementary in the mSUGRA parameter space, 
(iii) in NUHM SUGRA models, a discovery of 
$B(B_s \to \mu^+\mu^-) \simeq 5\times 10^{-9}$ at the LHC will cover
regions of the parameter space beyond the direct search 
for $pp \to b\phi^0 \to b\mu^+\mu^-$ with $L = 300$ fb$^{-1}$.

\end{abstract}

%

\maketitle

\section{Introduction}

In the minimal supersymmetric standard model (MSSM), the couplings of
down type quarks and leptons with neutral Higgs bosons are
proportional to $1/\cos\beta$. 
Thus a large value of $\tan\beta$ greatly enhances the discovery
potential for the Higgs decays into muon pairs with Higgs bosons 
produced in association with bottom quarks~\cite{Nikita,hbmm} 
as well as the branching fraction of the rare decay 
$B_s \to \mu^+\mu^-$ mediated by Higgs bosons~\cite{
Tata,bsmm,Carena,Ellis,Isidori}. 
In this article, we present results of Ref.~\cite{bsmm} for 
the discovery potential of the direct searches for the Higgs bosons 
$pp \to b\phi^0 \to b\mu^+\mu^- +X$ at the LHC and that of 
the indirect Higgs searches in $B_s \to \mu^+\mu^-$ 
at the Fermilab Tevatron Run II and the LHC within the framework of 
the minimal supersymmetric model (MSSM), the minimal supergravity 
unified model and supergravity models with non-universal Higgs masses 
at the unification scale.
The CERN Large Hadron Collider (LHC) has a great potential to discover 
both direct and indirect signals with muon pairs for neutral Higgs
bosons in supersymmetry models~\cite{bsmm}. 

\section{Minimal Supersymmetric Standard Model}
We evaluate the cross section of 
$pp \to b \phi^0 \to b \mu^+\mu^- +X, \phi^0 = h^0, H^0, A^0$ 
with the Higgs production cross section $\sigma(pp \to b \phi^0 +X)$ 
multiplied by the branching fraction of the Higgs decay into muon pairs 
$B(\phi^0 \to \mu^+\mu^-)$~\cite{hbmm} at the LHC with the 
factorization/renormalization scale $\mu_F = \mu_R = M_H/4$. 
We consider dominant physics backgrounds from $bg \to b \mu^+\mu^-$ 
as well as $gg \to b\bar{b}W^+W^-$ and $q\bar{q} \to b\bar{b}W^+W^-$
followed by the decays of $W^\pm \to \mu^\pm \nu_\mu$.
In addition, we have included the background from 
$pp \to b\mu^+\nu \mu^-\bar{\nu} +X$ and 
$pp \to j \mu^+\mu^- +X, j = g, q$ or $\bar{q}$ with $q = u, d, s, c$, 
where a jet is mistagged as a $b$ quark. 
We apply realistic cuts for an integrated luminosity ($L$) of 30 fb$^{-1}$ 
and for a higher integrated luminosity of 300 fb$^{-1}$. 
In our analysis for the MSSM, we adopt a common mass scale: 
\begin{eqnarray*}
M_{\rm SUSY} = m_{\tilde{g}} = m_{\tilde{f}} = \mu = -A_f \; .
\end{eqnarray*}

\begin{figure}[htb]
\centering\leavevmode
\epsfxsize=3.6in
\epsfbox{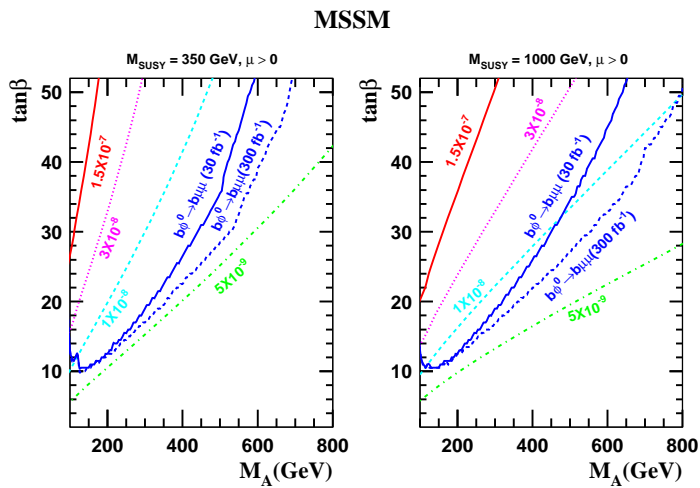}
\caption[]{
Discovery contours for $pp \to b\phi^0 \to b\mu\bar{\mu} +X$ at the
LHC and contours of the branching fraction of $B_s \to \mu^+\mu^-$ 
in the MSSM for 
(a) $m_{\rm SUSY} = 350$ GeV and (b) $m_{\rm SUSY} = 1000$ GeV.
The discovery region is the part of the parameter space above
the contour.
\label{fig:MSSM}}
\end{figure}

In Figure 1, we present contours for the branching fraction 
$B(B_s \to \mu^+\mu^-) = 1.5 \times 10^{-7}$ 
(current experimental limit~\cite{Bernhard:2005yn}), 
$3 \times 10^{-8}$, $1 \times 10^{-8}$, and 
$5 \times 10^{-9}$ as well as the discovery contours of 
$pp \to b\phi^0 \to b\mu^+\mu^- +X$ at the LHC.
For $M_{\rm SUSY} = 350$ GeV, LHC will be able to
discover $pp \to b\phi^0 \to b\mu^+\mu^- +X$ with an integrated
luminosity ($L$) of 30 fb$^{-1}$ in a significantly large region 
of the parameter space beyond $B(B_s \to \mu^+\mu^-) = 3\times 10^{-8}$. 
Furthermore, with a higher luminosity of 300 fb$^{-1}$, the LHC
discovery contour for direct Higgs search with 
$M_{\rm SUSY} = 1000$ GeV is very close to the contour for 
$B(B_s \to \mu^+\mu^-) = 5 \times 10^{-9}$.

\section{Minimal Supergracity Unified Model}
In the minimal supergravity (mSUGRA) model, supersymmetry (SUSY) is 
broken in a hidden sector and SUSY breaking is communicated to 
the observable sector through gravitational interactions.  
The mSUGRA parameters are chosen to be a common scalar mass ($m_0$),  
a common gaugino mass ($m_{1/2}$), a common trilinear coupling ($A_0$), 
sign of the Higgs mixing parameter ($\mu$), and  
the ratio of Higgs field vacuum expectation values at the electroweak  
scale ($\tan\beta \equiv v_2/v_1$).

\begin{figure}[htb]
\centering\leavevmode
\epsfxsize=3.8in
\epsfbox{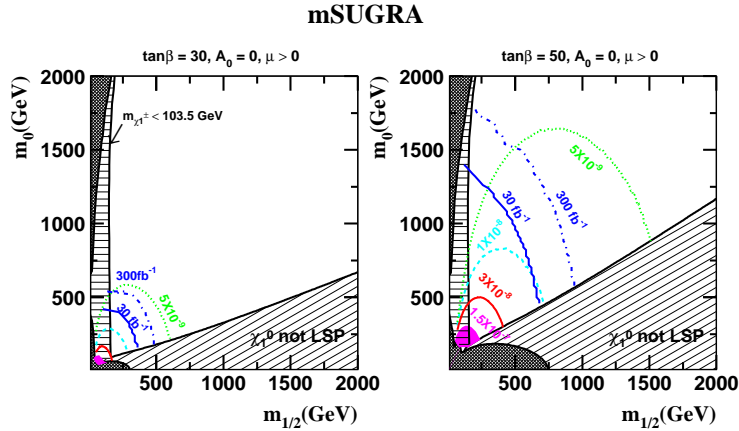}
\caption[]{
Discovery contours for $pp \to b\phi^0 \to b\mu\bar{\mu} +X$ 
at the LHC and contours of the branching fraction of $B_s \to \mu^+\mu^-$ 
in the minimal supergravity unified model for 
(a) $\tan\beta = 30$ and (b) $\tan\beta = 50$. 
Also shown are the parts of the parameter space
(i)~excluded by theoretical requirements (slant-hatched and dark shaded),
or (ii)~excluded by the chargino search at LEP 2 (horizontally-hatched).
\label{fig:MSUGRA}}
\end{figure}

Figure 2 displays the discovery contours of 
$pp \to b\phi^0 \to b\mu^+\mu^- +X$ 
for an integrated luminosity of 30 fb$^{-1}$ and 300 fb$^{-1}$ 
at the LHC as well as contours for the branching fraction of  
$B(B_s \to \mu^+\mu^-)$ in the $(m_{1/2},m_0)$ plane 
of the mSUGRA model with $\tan\beta =$ 30 and 50. 
If $\tan\beta \leq 30$, only a tiny region with small values of
$m_{1/2}$ and $m_0$ will likely lead to observable signals for either 
$B_s \to \mu^+\mu^-$ at the Tevatron Run II or 
$pp \to b\phi^0 \to b\mu^+\mu^- +X$ at the LHC. 
For $\tan\beta \leq 40$, direct searches for 
$pp \to b\phi^0 \to b\mu^+\mu^- +X$ at the LHC with $L = 30$ fb$^{-1}$ 
covers a much larger region in the mSUGRA parameter space than 
$B(B_s \to \mu^+\mu^-) \ge 1\times 10^{-8}$.
If $\tan\beta \geq 50$, both discovery channels at the LHC become 
complementary. The direct searches for 
$pp \to b\phi^0 \to b\mu^+\mu^- +X$ with $L = 300$ fb$^{-1}$ 
covers a significant region beyond the contour of 
$B(B_s \to \mu^+\mu^-) = 5\times 10^{-9}$. 
Likewise, the rare decay with 
$B(B_s \to \mu^+\mu^-) \ge 5\times 10^{-9}$ 
covers a large region beyond the discovery contour of the direct search for 
$b\phi^0 \to b\mu^+\mu^-$ with $L = 300$ fb$^{-1}$.

\section{mSUGRA with Non-universal Higgs Masses}
In our analysis for supergravity models with non-universal Higgs
masses at the unification scale (NUHM SUGRA models), 
the GUT-scale Higgs masses are parametrized as: 
$m_{H_i}^2 ({\rm GUT}) = (1+\delta_i) m_0^2$ with $i = 1,2$.
We find that a decrease in $m_{H_1}$ with a negative $\delta_1$ 
as well as an increase in $m_{H_2}$ with a positive $\delta_2$ 
at $M_{\rm GUT}$ will lead to a smaller mass at the electroweak scale 
for the Higgs pseudoscalar ($A^0$) or the heavier Higgs scalar ($H^0$) 
than that in the mSUGRA model. We choose three sets of
values for $\delta_i$ to study the Higgs discovery potential: 
(i) $\delta_1 = -0.5, \delta_2 = 0$,
(ii) $\delta_1 = 0, \delta_2 = 0.5$, and
(iii) $\delta_1 = -0.5, \delta_2 = 0.5$.

\begin{figure}[htb]
\centering\leavevmode
\epsfxsize=3.8in
\epsfbox{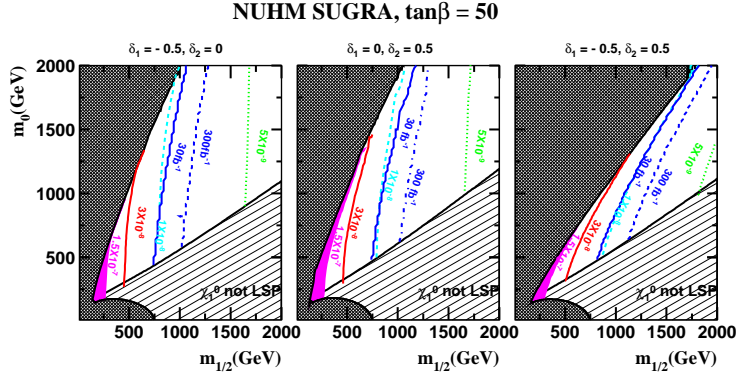}
\caption[]{
The $5\sigma$ contours for $pp \to b\phi^0 \to b\mu\bar{\mu} +X$ 
at the LHC with $L = 30$ fb$^{-1}$ and 300 fb$^{-1}$ 
as well as contours for the branching fraction of $B_s \to \mu^+\mu^-$ 
in the ($m_{1/2},m_0$) plane of a NUHM SUGRA model
with $\tan\beta = 50, \mu >0$, $A_0 = 0$ and non-universal boundary conditions
(a) $\delta_1 = -0.5$ and $\delta_2 = 0$, 
(b) $\delta_1 = 0$ and $\delta_2 = 0.5$, 
(c) $\delta_1 = -0.5$ and $\delta_2 = 0.5$.
\label{fig:NONU-SUGRA}
}
\end{figure}

In Figure 3, we present the discovery contours of $b\phi^0 \to b\mu^+\mu^-$ 
for integrated luminosities of 30 fb$^{-1}$ and 300 fb$^{-1}$ at the
LHC as well as contours for branching fraction 
$B(B_s \to \mu^+\mu^-$ in the $(m_{1/2},m_0)$ plane with $\tan\beta = 50$. 
In all three NUHM SUGRA cases, $m_A$ and $m_H$ are smaller than those 
in the mSUGRA model for the same values of $m_0$ and $m_{1/2}$. 
Consequently, both $b\phi^0 \to b\mu^+\mu^-$ and $B_s \to \mu^+\mu^-$ 
will be able to cover larger regions in the $(m_{1/2},m_0)$ plane.
We note that for $\tan\beta \geq 50$, 
the observable region for $b\phi^0 \to b\mu^+\mu^-$ at the LHC with 
$L = 30$ fb$^{-1}$ is comparable to that of 
$B(B_s \to \mu^+\mu^-) \ge 1\times 10^{-8}$. 
If $m_{H_2}$ is larger than $m_0$ with $\delta_2 = 0.5$, 
the theoretically disfavored region grows rapidly as the value of
$\tan\beta$ increases. 
If both Higgs boson masses are different from the common scalar mass 
at $M_{\rm GUT}$, then theoretically favored region shrinks greatly.

In supersymmetric models, the muon pair discovery channels offer 
great promise to detect Higgs signatures in $B_s \to \mu^+mu^-$ 
as well as in $pp \to b\phi^0 \to b\mu^+\mu^- +X$ at the CERN LHC. 
In mSUGRA models, the branching fraction of $B_s \to \mu^+\mu^-$ and 
the significance of $pp \to b\phi^0 \to b \mu^+\mu^-+X$ 
are greatly improved by a large $\tan\beta$ because  
the large $b\bar{b}\phi^0$ couplings make $m_A$ and $m_H$ small 
and enhance the Higgs bosons production.
In both mSUGRA model and NUHM SUGRA models, the direct signal of 
$b\phi^0 \to b\mu^+\mu^-$ at the LHC can be discovered with a 
luminosity of 30 fb$^{-1}$ in a large space that is comparable to 
that of $B(B_s \to \mu^+\mu^-) = 1 \times 10^{-8}$ for $\tan\beta \leq
50$. 

If $\tan\beta \geq 50$, the regions covered by 
$B(B_s \to \mu^+\mu^-) \ge 5\times 10^{-9}$ and the discovery
region for $b\phi^0 \to b\mu^+\mu^-$ with 300 fb$^{-1}$ are
complementary in the mSUGRA parameter space.
However, in NUHM SUGRA models, a discovery of 
$B(B_s \to \mu^+\mu^-) \simeq 5\times 10^{-9}$ at the LHC will cover
regions of the parameter space beyond the direct search 
for $b\phi^0 \to b\mu^+\mu^-$ with $L = 300$~fb$^{-1}$.

\section*{Acknowledgments}

This research was supported in part by the U.S. Department of Energy
under Grants No.~DE-FG02-04ER41305 and No.~DE-FG02-03ER46040.

%

\end{document}